# The Influence of Boreal Summer Madden-Julian Oscillation on Precipitation Extremes in Indonesia

Fadhlil R. Muhammad[1], Sandro W. Lubis[2], Sonni Setiawan[3]

[1]School of Geography, Earth, and Atmospheric Sciences, The University of Melbourne, Australia
[2]Rice University, Houston, USA
[3]Department of Geophysics and Meteorology, IPB University

`fadhlilrizki@student.unimelb.edu.au`

**Abstract.** This study examines the influence of Madden-Julian Oscillation (MJO) on Indonesian precipitation during the extended boreal summer (May – September). The MJO is one of the dominant intra-seasonal variabilities that influence the extreme precipitation in the tropics, especially in Indonesia. The episodes of intense precipitation (95th percentiles) during active MJO phases from 1998-2015 are evaluated using the daily precipitation datasets from the gridded Asian Precipitation–Highly Resolved Observational Data Integration Towards Evaluation of Water Resources (APHRODITE) and several station rain gauges. The boreal summer MJO influences the probability of extreme precipitation, especially in the west and north parts of Indonesia. The west part experiences an increase in the probability of extreme precipitation by up to 55% and 80% during phases 2 and 3, respectively. Moreover, the extreme precipitation probability in the north part increases by up to 40 - 70% during phases 2 - 4. On the other hand, the influence of MJO is relatively weak in the south and east parts of Indonesia. The contrast of the precipitation response between the north and south parts of Indonesia is consistent with the northward movement of boreal summer MJO.

## 1 Introduction

The Madden-Julian Oscillation is one of the most dominant modes of variability in the tropics on the intraseasonal timescale. The MJO is characterized by the eastward propagation of large cloud clusters over the tropical region with a period of 30 – 60 days [1–4]. A typical case of MJO is comprised of a pair of wet and dry phases associated with enhanced and suppressed convection, respectively [3]. The propagation of MJO can be divided into eight phases, with each phase has a period of 4 – 8 days [5]. The MJO propagation can be described as follows:

- Phase 1 is located over the western Indian Ocean and Africa.
- Phases 2 and 3 are located over the Indian Ocean.
- Phases 4 and 5 are located over the Maritime Continent, including Indonesia.
- Phases 6 and 7 are located over the western Pacific.
- Finally, the MJO ends its propagation in phase 8 over the Pacific Ocean.

The MJO can enhance or suppress the convection over the region, depending on its phase and region. Over the Maritime Continent, including Indonesia, a typical boreal winter MJO enhances the convection during phases 2 – 4 and suppresses the convection during phases 6 – 8 [6, 7]. This enhanced convection results in an increase in precipitation over the region. Furthermore, it has also been shown that the combination between large-scale convection, moisture convergence, and upward moisture advection associated with the boreal winter MJO can increase the probability of extreme precipitation events over Indonesia by up to 70% [7].

While the boreal winter MJO has been extensively studied, its counterpart, the boreal summer MJO, is still insufficiently researched. The main difference between boreal winter and boreal summer MJO is its propagation [3, 8]. The boreal winter MJO often thought of as the "dominant" MJO, exhibits a more southward tilt of propagation when it passes the Maritime Continent. On the other hand, the boreal summer MJO shows more northward tilt of the propagation when passing the Maritime Continent.

Because of the difference in propagation, the impact of boreal winter and summer MJO should be different, especially over Indonesia. Therefore, in this study, we will investigate these three main questions:

1. What are the impacts of boreal summer MJO on precipitation extremes in Indonesia?
2. Does the impact differ from boreal winter MJO? If so, in what location does it differs?
3. What are the underlying dynamics behind the impact?

## 2    Data and Methods

In this study, we focus on the extended boreal summer season (May – September). We use daily gridded precipitation datasets from Asian Precipitation–Highly Resolved Observational Data Integration Towards Evaluation of Water Resources (APHRODITE) with the period of 1998 – 2015 and a spatial resolution of 0.25° × 0.25° [9]. Furthermore, we also use 63 high-quality daily rain-gauge data from BMKG from 1987 to 2017.

To understand the dynamics behind the impact, we use bandpass-filtered daily anomalies with frequency cut-offs of 20 – 100 days [3]. Then, we calculate anomaly composites for each phase of the MJO for several variables such as outgoing longwave radiation (OLR), vertically integrated moisture flux convergence (VIMFC), and vertical moisture advection. The VIMFC and vertical moisture advection are calculated using relative humidity, air temperature, zonal, meridional, and vertical wind at various pressure levels (i.e., 1000 – 100 hPa) [7]. The OLR datasets are obtained from NOAA and are used as a proxy for convection [10]. The other variables are obtained from NCEP-DOE Reanalysis II [11]. Both datasets consist of daily data from 1998 to 2015 with a spatial resolution of 2.5° × 2.5°.

This study defines the boreal summer MJO using the Real-time Multivariate MJO Index (RMM) [5]. Before the analysis, the index is selected, so it only contains data from May to August. Then, the days with an MJO index of more than one are categorized as active MJO events, while those with less than one are categorized as weak MJO events.

We use probability composites to investigate the changes in the probability of extreme precipitation [7, 12]. The probability of precipitation exceeding the threshold (95$^{th}$ percentile, Fig. 1) is calculated for each active phase of the MJO. Then, the changes in the probability can be written as follows [7]:

$$\Delta P_{MJO} = \frac{P_{MJO}(x \geq x_c) - P_{summer}(x \geq x_c)}{P_{summer}(x \geq x_c)} \times 100\% \quad (1)$$

Where $\Delta P$ is the percentage changes in the precipitation extremes probability, $P_{MJO}$ is the probability of non-zero precipitation exceeding the threshold ($x_c$) for each active phase of the MJO, and $P_{summer}$ is the same as $P_{MJO}$, except for all boreal summer days. The $\Delta P$ is calculated at each grid and station. Moreover, we also calculate the area average of $\Delta P$ for each region in Indonesia (Fig.1). The significance of anomaly and probability composites in this study are tested using the bootstrap test method with 1,000 synthetic composites [13].

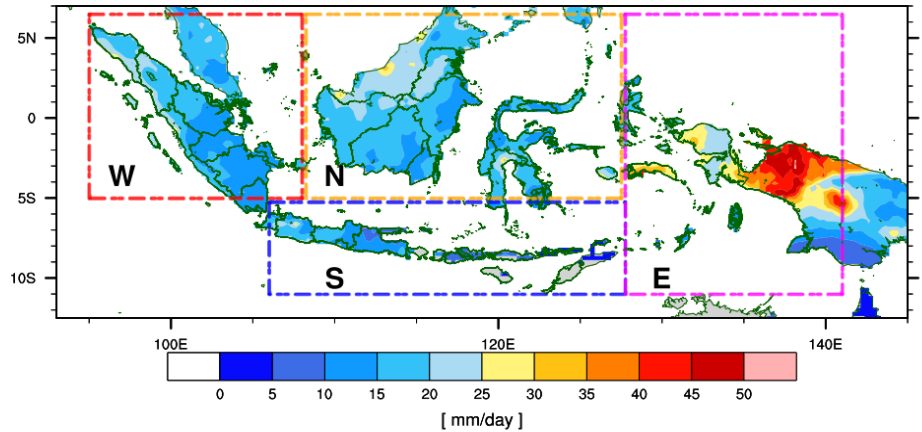

**Fig. 1.** The 95th percentile of precipitation during boreal summer (May - September). Red, orange, blue, and pink boxes represent the western, northern, southern, and eastern parts of Indonesia, respectively. Grey shades denote missing values.

## 3 Results and Discussion

### 3.1 MJO impact on regional precipitation anomalies

First, we analyze the impact of boreal summer MJO on the precipitation anomalies over Indonesia. Figure 2 shows the filtered anomaly composites during each phase of the MJO. The increase/decrease of precipitation occurs mainly over Sumatra, Borneo, Sulawesi, Molucca, and west of Papua islands by up to 3 mm/day. In general, the results show that boreal summer MJO increases precipitation over the Indonesian region during phases 2 – 3 and decreases precipitation during phases 6 – 8.

From this result, we found that the impact on precipitation differs from that of boreal winter MJO. During boreal winter MJO, the precipitation affects most of the region in Indonesia, including Java [7].On the other hand, the boreal summer MJO influences mostly on the northern region of Indonesia, with a very small influence over the southern region (i.e., Java, Bali, and Nusa Tenggara islands).

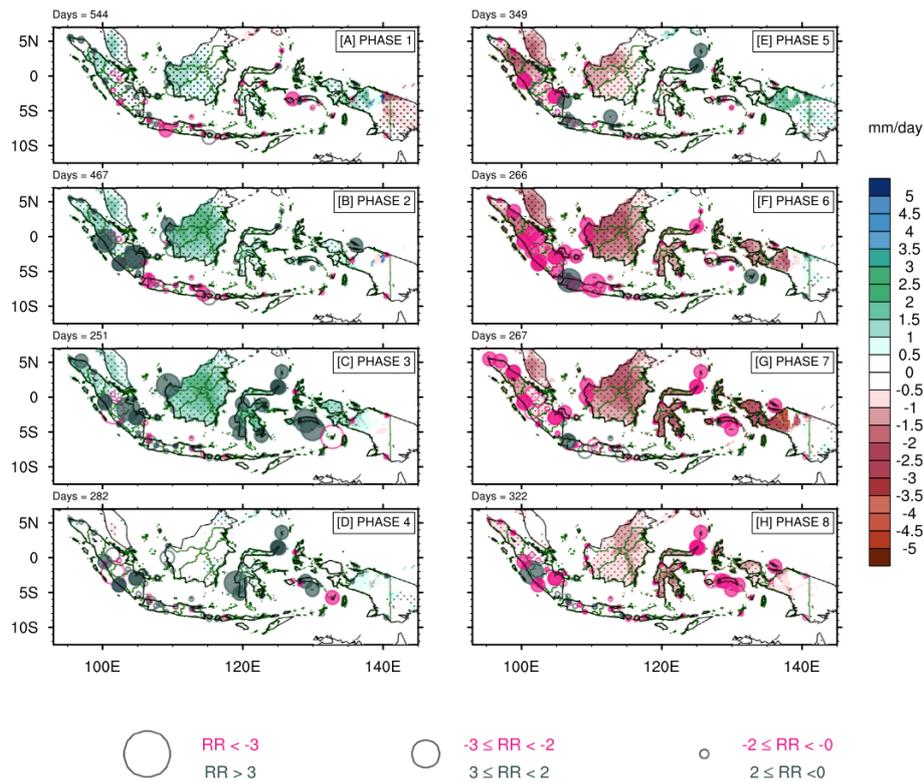

**Fig. 2.** Composites of precipitation anomaly for each MJO phase (A - H) during the boreal summer (May - September) were observed by the rain-gauges (circle) and APHRODITE (shading). The filled circle and dots indicate the values that exceeded the 95% confidence level for the rain-gauges and APHRODITE, respectively.

## 3.2 MJO impact on precipitation extremes

In this section, we investigate the impact on the precipitation extremes. Figure 3 shows the changes in the probability of precipitation that exceeds the 95$^{th}$ percentile threshold (Fig. 1). Based on the figure, we conclude that MJO increases the probability of precipitation extremes in Indonesia during phases 2 – 3 and decreases the probability during phases 6 – 8. The increase in the probability is around 60 – 150% over the Sumatra, Borneo, and Sulawesi islands. Alternatively, during phases 6 – 8, the MJO decreases the probability by around 30 – 60%.

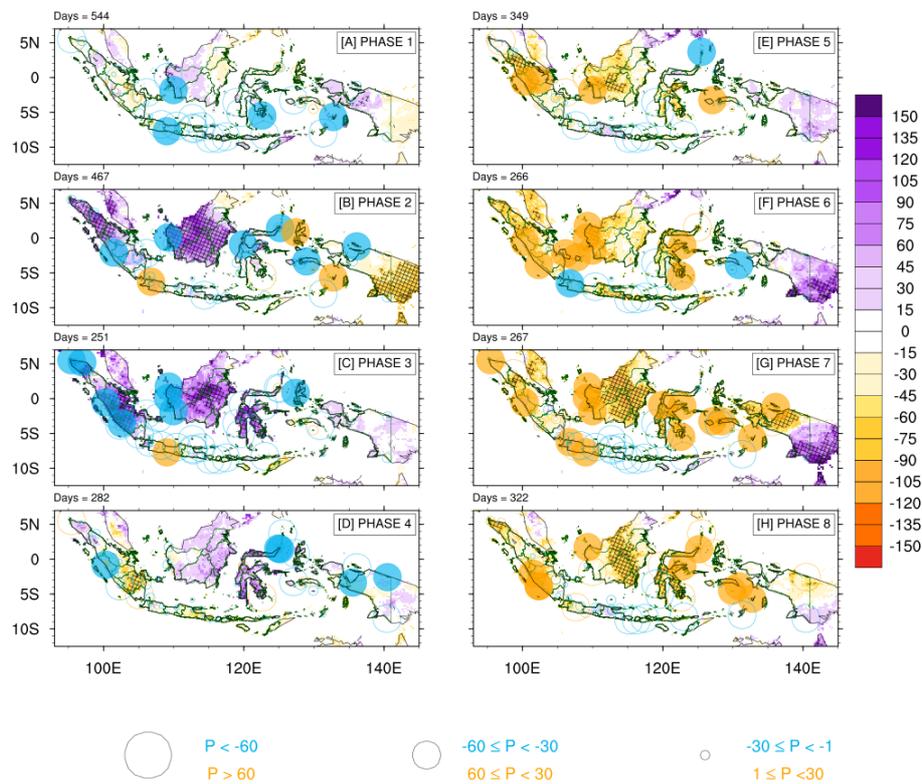

**Fig. 3.** Percentage changes in the probability of extreme precipitation events observed by the rain-gauges (circle) and APHRODITE (shading) data for each phase of the MJO (A-H). Filled circle and hatching indicate the values that exceeded the 95% confidence level for the rain-gauges and APHRODITE, respectively.

The strongest impact is seen during phase 3 of the MJO over central Borneo, central Sumatra, and most Sulawesi (Fig. 3). During this phase, the probability of precipitation extremes increased by up to 150%. The rain-gauge observation corroborates the result, showing a high probability increase (>60%), particularly over Sumatra and Borneo. In contrast, a very small impact is found over the southern region of Indonesia. The difference between APHRODITE and rain-gauge observation could be due to the local effects (i.e., topography and diurnal cycle) [14–16]. Nevertheless, This result is generally consistent with the northward tilt of the boreal summer MJO propagation.

    To better understand the regional differences of the impact, we calculate the area average of the impact based on geographical locations as in Fig 1. Figure 4 shows that the boreal summer MJO increases the probability of precipitation extremes over Indonesia's western and northern regions (Fig 4a and 4b). Over the western region, the MJO increases the probability by around 50% during phase 2 and 85% during phase 3. While over the northern region, the MJO increases the probability by approximately 55%, 75%, and 35% during phases 2, 3, and 4, respectively. The eastern region of Indonesia is also affected by the boreal summer MJO, albeit in a smaller magnitude. The MJO increases the probability of precipitation extremes over the eastern region by around 20 – 30% during phases 3 and 4. On the contrary, the impact on the southern region is very small, only around up to 10% of the increase in the probability.

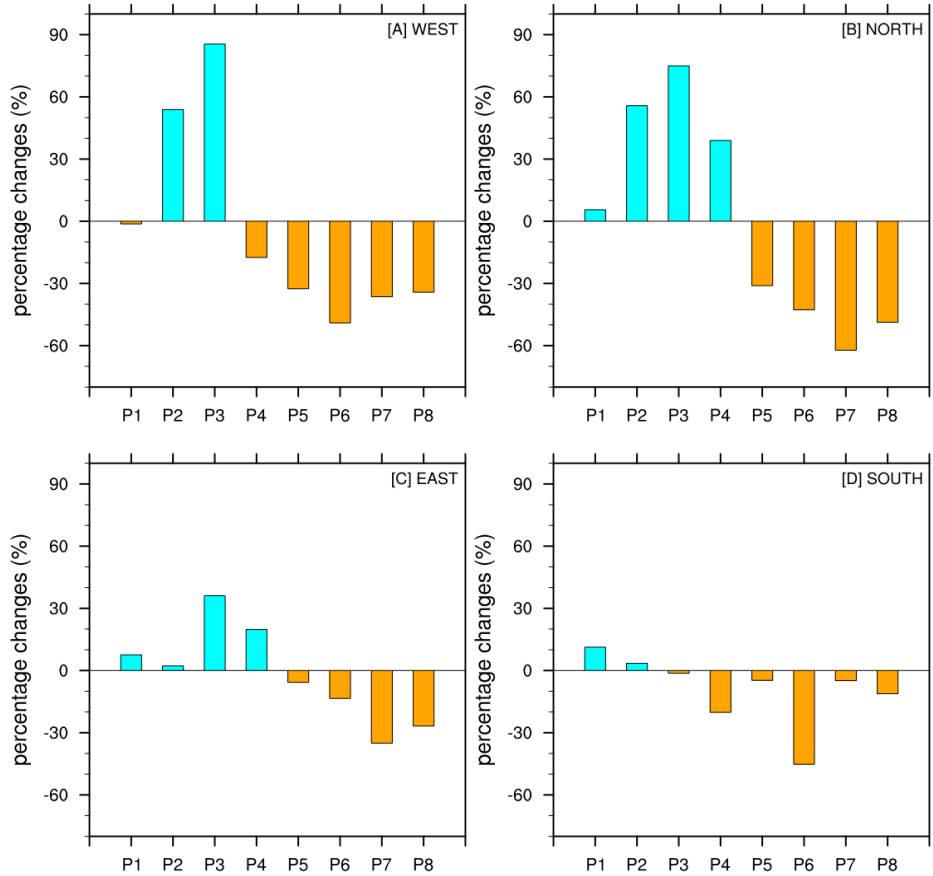

**Fig. 4.** Area-average of the probability composites for different regions in Indonesia.

In general, the impact of the boreal summer MJO on precipitation extremes is the strongest during phases 2 – 3, affecting most of the western and northern part of Indonesia, similar to that of precipitation anomalies. The location of the impact differs from that of boreal winter MJO. The boreal winter MJO increases the probability of precipitation extremes over most of the regions of Indonesia [7], while the boreal summer MJO only strongly impacts the western and northern regions of Indonesia.

### 3.3  The dynamical links between the boreal summer MJO and precipitation extremes

There are two important factors that are responsible for the increase in the probability of precipitation extremes, specifically the VIMFC [7, 17, 18] and vertical moisture advection [7, 19–21]. In this section, we investigate the role of VIMFC and vertical moisture advection anomalies associated with MJO on the precipitation extremes. The VIMFC is calculated as follows [7, 22]:

$$\text{VIMFC} = -\frac{1}{g} \int_{1000\text{ hPa}}^{100\text{ hPa}} \left[-\left(\vec{\nabla} \cdot (q\mathbf{u}_H)\right)\right] dp \qquad (2)$$

Where g is gravity (9.8 ms-2), $\mathbf{u}_H$ is the horizontal wind, and q is specific humidity (kg/kg). Additionally, the vertical moisture advection can be written as:

$$\text{vert. qadv} = -\omega \frac{\partial q}{\partial p} \qquad (3)$$

Where ω is the vertical component of wind and p is pressure level. A large-scale moisture convergence from 1000 – 100 hPa and vertical moisture advection due to boreal winter MJO triggers the occurrence of precipitation extremes in Indonesia [7].

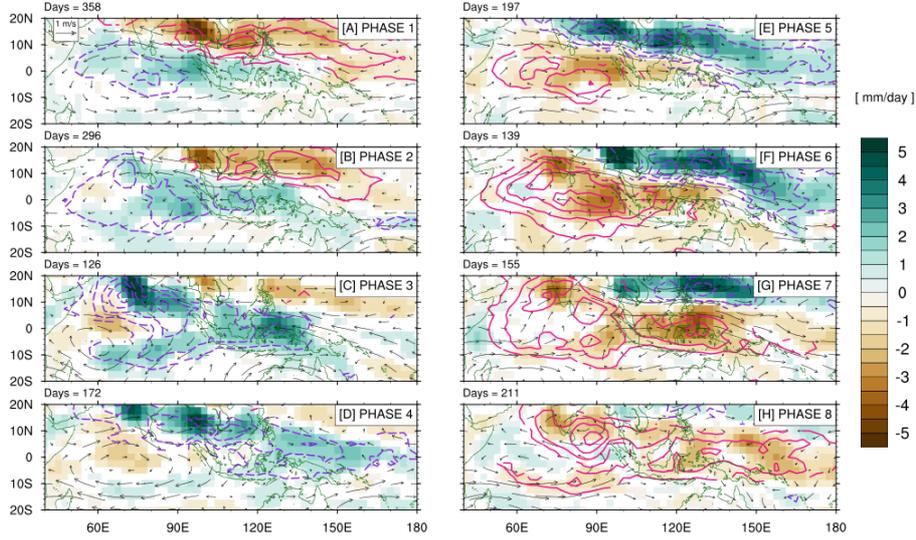

**Fig. 5.** Anomaly composites of VIMFC (shading, mm day$^{-1}$), OLR (contour, Wm$^{-2}$), and 850-hPa wind (ms$^{-1}$). Dashed blue (Solid Red) contour denotes the negative (positive) anomaly of OLR. The contour is from -50 to 50 with an interval of 5. The values shown are significant at a 95% confidence level.

Figure 5 shows the filtered anomaly composites of VIMFC and OLR. Based on the figure, during phase 2, the MJO activity enhances the convection over most of Indonesia's western and northern regions (Fig 5b). This enhanced convection is collocated with enhanced moisture convergence. Furthermore, during phase 3, the enhanced convection and convergence due to MJO are getting stronger. This result is consistent with the increase of precipitation extremes' probability. As expected, the boreal summer MJO propagation is more tilted to the north instead of to the south like boreal winter MJO. However, the northward tilt is not observed until after phase 4. This is consistent with the boreal summer MJO's impact on precipitation (Fig 3 and 4), which only affected the northern part of Indonesia. In contrast, the MJO activity suppresses the convection and moisture convergence over most of the Indonesian regions during phases 6 – 8, resulting in a decrease in the probability of precipitation extremes.

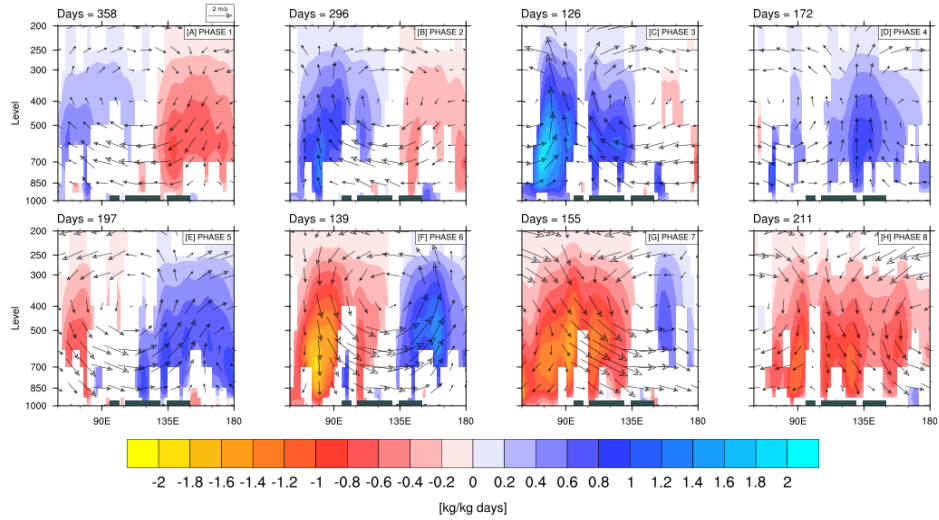

**Fig. 6.** Anomaly composites of vertical moisture advection (shading) and wind (vector) averaged over 11°N – 11°S for each phase of MJO (A-H). Shaded values are significant at a 95% confidence level.

The enhanced (suppressed) convection and moisture convergence are consistent with the upward (downward) moisture advection due to MJO. The upward moisture advection is collocated with the enhanced convergence, resulting in an increase in precipitation and the precipitation extremes probability. Similarly, the downward moisture advection is collocated with the suppressed convergence, causing a decrease in precipitation and the probability of the extremes. The strongest upward moisture advection is observed during phase 3 over 70° – 90°E, particularly over the 850 – 300 hPa, which is consistent with the strongest impact on precipitation. Moreover, throughout all phases, the upward moisture advection is mostly seen at the middle to the upper-level troposphere (i.e., 700 – 300 hPa). While the result is generally similar to that of boreal winter MJO, the boreal summer MJO exhibits less visible lower level (i.e., 1000 – 850 hPa) upward moisture advection compared to that of boreal winter MJO [7]. The lower-level advection favors a suitable environment for convection, resulting in a heavier rainfall [7, 23, 24].

    In summary, the precipitation anomaly and the probability of precipitation extremes are directly associated with the VIMFC and vertical moisture advection due to MJO. The enhanced (suppressed) moisture convergence and upward (downward) moisture advection contribute strongly to the increased (decreased) precipitation and the probability of precipitation extremes in Indonesia.

## 4  Conclusions

The impact of boreal summer MJO on precipitation extremes and its dynamical links are investigated. We found that the boreal summer MJO significantly modulates the precipitation in Indonesia, particularly during phases 2 and 3. To sum up, we conclude the results as follows:

1. The impact of the boreal summer MJO is the strongest over the western and northern parts of Indonesia, with moderate impact over the eastern region. In contrast, the impact on the southern region is very small due to the northward tilt of boreal summer MJO propagation.
2. Phase 2 and 3 of the boreal summer MJO have the strongest impact on the Indonesian region. Both phases can increase the probability of precipitation extremes by up to 150% over the western and northern parts and around 30% over the eastern part of Indonesia.
3. On the contrary, the suppressed phase of the boreal summer MJO (phases 6 – 8) affects most of the Indonesian region, including the southern part. The suppressed phase of the MJO decreases the precipitation and the probability of precipitation extremes in Indonesia.
4. The enhanced moisture convergence and vertical moisture advection are directly responsible for the increase in precipitation and extreme precipitation events.

To sum up, the results suggest that MJO has a considerable impact on precipitation extremes during boreal summer in Indonesia, specifically over the western and northern parts of Indonesia. While the impact of boreal summer MJO is not as widespread as its boreal winter counterparts, the percentage changes of precipitation extremes' probability are high, reaching up to 150%. Moreover, due to the widespread decrease of precipitation during MJO's suppressed phase, future study is still required to better understand this response, which may be related to the complex interaction between MJO and other types of tropical disturbances, such as equatorial waves [25–27], Indian Ocean Dipole [28] and summer monsoon (also known as BSISO) [29, 30] in modulating summertime rainfall variability over Indonesia. Overall, this research highlights the importance of the MJO in providing an important source for probabilistic prediction of summertime precipitation extremes in Indonesia.